\RequirePackage{fix-cm}
\documentclass[twocolumn]{svjour3}          
\usepackage{graphicx}
\journalname{Experimental Astronomy, Springer,}
\begin{document}

\title{Photometric redshift estimation based on data mining with PhotoRApToR}



\author{S. Cavuoti, M. Brescia V. De Stefano, G. Longo
}

\authorrunning{Cavuoti et al.} 

\institute{S. Cavuoti \at
              INAF - Astronomical Observatory of Capodimonte, Italy \\
              Tel.: +39-081-5575553\\
              Fax: +39-081-456710
              \email{cavuoti@na.astro.it}           
           \and
           M. Brescia \at
              INAF - Astronomical Observatory of Capodimonte, Italy
           \and
           V. De Stefano, G. Longo \at
              Dep. of Physics - University Federico II of Naples, Italy \\
}

\date{Received: Sep 21, 2014 / Accepted: Jan 24, 2015}

\maketitle

\begin{abstract}
Photometric redshifts (photo-z) are crucial to the scientific exploitation of modern panchromatic digital surveys.
In this paper we present PhotoRApToR (Photometric Research Application To Redshift): a Java/C++ based desktop application
capable to solve non-linear regression and multi-variate classification problems, in particular specialized for photo-z estimation. It embeds a machine
learning algorithm, namely a multi-layer neural network trained by the Quasi Newton learning rule, and special tools dedicated to pre- and post-processing data.
PhotoRApToR has been successfully tested on several scientific cases. The application is available for free download
from the DAME Program web site.

\end{abstract}

\section{Introduction}
The ever growing amount of astronomical data provided by the new large scale digital surveys in a wide range of wavelengths
of the electromagnetic spectrum has been challenging the way astronomers carry out their everyday analysis of astronomical sources
and we can safely assert that the human ability to directly  visualize and correlate astronomical data has been pushed to its limits in
the past few years. As a consequence of the fact that data have become too complex to be effectively
managed and analysed with traditional tools, a new methodological shift is emerging and Data Mining (DM) techniques are becoming
more and more popular in tackling knowledge discovery problems. A typical problem which is addressed with these new techniques is that
of the evaluation of photometric redshifts.
The request for accurate photometric redshifts (photo-z) has increased over the years due both to the advent of a
new generation of multi-band surveys (see for example Connolly et al. 1995, \cite{connolly1995}) and to the availability of large public
datasets which allowed to pursue a wide variety of scientific cases. Ongoing and future large-field
public imaging projects, such as Pan-STARRS (Farrow et al. 2014, \cite{farrow2014}), KiDS\footnote{http://www.astro-wise.org/projects/KIDS/} (Kilo-Degree Survey),
DES (Dark Energy Survey, \cite{des2005}), the planned surveys with LSST (Large Synoptic Survey Telescope, Ivezic et al. 2009, \cite{ivezic2009}) and Euclid (Red Book, \cite{euclidredbook}), rely on accurate photo-z to achieve their scientific goals.

Photo-z are in fact essential in constraining dark matter and dark energy through weak gravitational lensing (Serjeant 2014, \cite{serjeant2014}), for the identification of galaxy clusters and groups (e.g. Capozzi et al. 2009, \cite{capozzi2009}), for type Ia Supernovae, and for the study of the mass function of galaxy clusters (Albrecht et al. 2006, \cite{albrecht2006}, Peacock et al. 2006, \cite{peacock2006},
and Umetsu et al. 2012, \cite{umetsu2012}), just to quote a few applications.
Photometric filters integrate fluxes over a quite large interval of wavelengths and, therefore,
the accuracy of photometric redshift reconstruction is worse than that of
spectroscopic redshifts. On the other hand, in the absence of the minimal telescope time necessary to determine spectroscopically
the redshifts for all sources in a sample, photometric redshifts methods provide a much more convenient way
to estimate the distance of such sources.
The physical mechanism responsible for the correlation existing between the photometric features and
the redshift of an astronomical source, is the change in the observed fluxes caused by the fact that,
due to the stretch introduced by the redshift, prominent features of the spectrum move across the
different filters of a photometric system.

This mechanism implies a non-linear mapping between the photometric parameter space of the galaxies
and the redshift values. This non linear mapping function can be inferred using advanced statistical and data mining methods in order to evaluate photometric
estimates of the redshift for a large number of sources.
All existing implementations can be broadly categorized into two classes of methods: theoretical and empirical.
\textit{Theoretical methods} use template based Spectral Energy Distributions (SEDs), obtained either from observed galaxy spectra
or from synthetic models.
These methods require an extensive a-priori knowledge about the physical properties of the objects, hence they may be biased by such
information. They, however, represent the only viable method when dealing with faint objects outside the spectroscopic limit (Hildebrandt et al. 2010, \cite{hildebrandt2010} and references therein).

When accurate and multi-band photometry for a large number of objects is complemented by spectroscopic redshifts for a
statistically significant sub-sample of the same objects, \textit{empirical methods} might offer greater accuracy.
This sample needs, however, to be statistically representative of the parent population.
The spectroscopic redshifts of this sub-sample are then used to constrain the fit of an interpolating function mapping the
photometric parameter space. Different methods differ mainly in the way such interpolation is performed.

From the data mining point of view, the evaluation of photo-z is a supervised learning problem (Tagliaferri et al. 2002, \cite{tagliaferri2002}), (Hildebrandt et al. 2010, \cite{hildebrandt2010}, where a set of \textit{examples} is used by the method to learn how to reconstruct the relation between the \textit{parameters} and the \textit{target} (Brescia 2012, \cite{brescia2012a}).
In the specific case of photometric redshifts, the parameters are fluxes, magnitudes or colors of the sources while the
targets are the spectroscopic redshifts.

A con of this approach being that, as it happens for all interpolative problems, such methods may suffer to extrapolate and therefore
they are effective only when applied to galaxies with photometry that lie within the range of fluxes/magnitudes and redshifts well sampled by the
training set.
In this paper we present PhotoRApToR (Photometric Research Application To Redshift), namely a Java based desktop application capable
to solve regression and classification problems which has been finely tuned for photo-z estimation.
It embeds a Machine Learning (ML) algorithm, in the specific case a particular instance of a multi-layer neural network, and special tools
dedicated to pre- and post-processing data. The machine learning model is the MLPQNA (Multi Layer Perceptron trained by the Quasi Newton
Algorithm), which has proven to be particularly powerful photo-z estimator, also in presence of relatively small spectroscopic Knowledge
Base (KB) (Cavuoti et al. 2012, \cite{cavuoti2012}), (Brescia et al. 2013, \cite{brescia2013}).
The application is available for download from the DAME program web site\footnote{http://dame.dsf.unina.it/dame\_photoz.html\#photoraptor}.
This paper is organized as follows: in Sect.~\ref{photoraptor} we describe the Java application; in Sect.~\ref{photoz} we discuss in
some details how the evaluation of photometric redshifts is performed. Sect.~\ref{others} describe other functionalities provided by the application, while Sect.~\ref{comparison} is dedicated to a comparison between PhotoRApToR and an alternative public machine learning tool. Finally in Sect.~\ref{conclusions} we outline some lessons which
were learned during the implementation of PhotoRaPToR and draw some future developments.

\section{PhotoRApToR}
\label{photoraptor}
\begin{figure*}
\centering
\includegraphics[width=12cm]{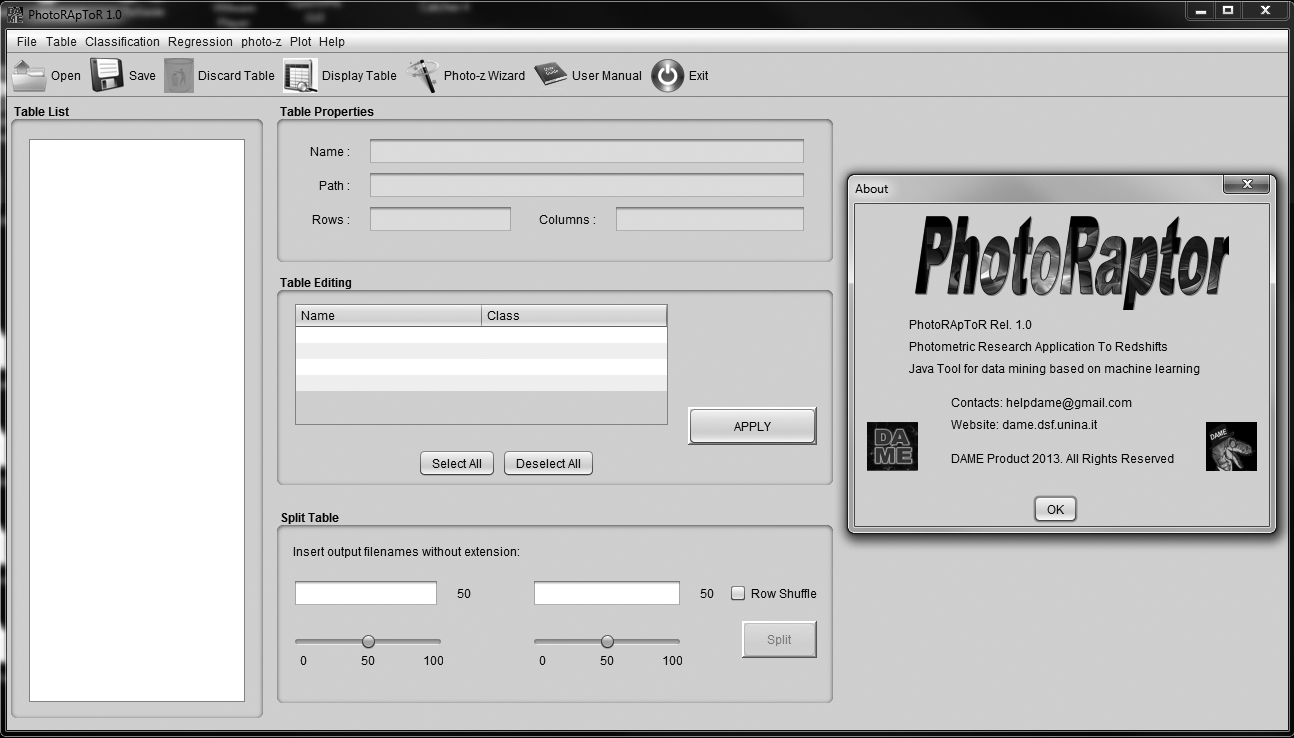}
\caption{The PhotoRApToR main window.}
\label{fig:main}
\end{figure*}

\begin{figure*}
\centering
\includegraphics[width=9cm]{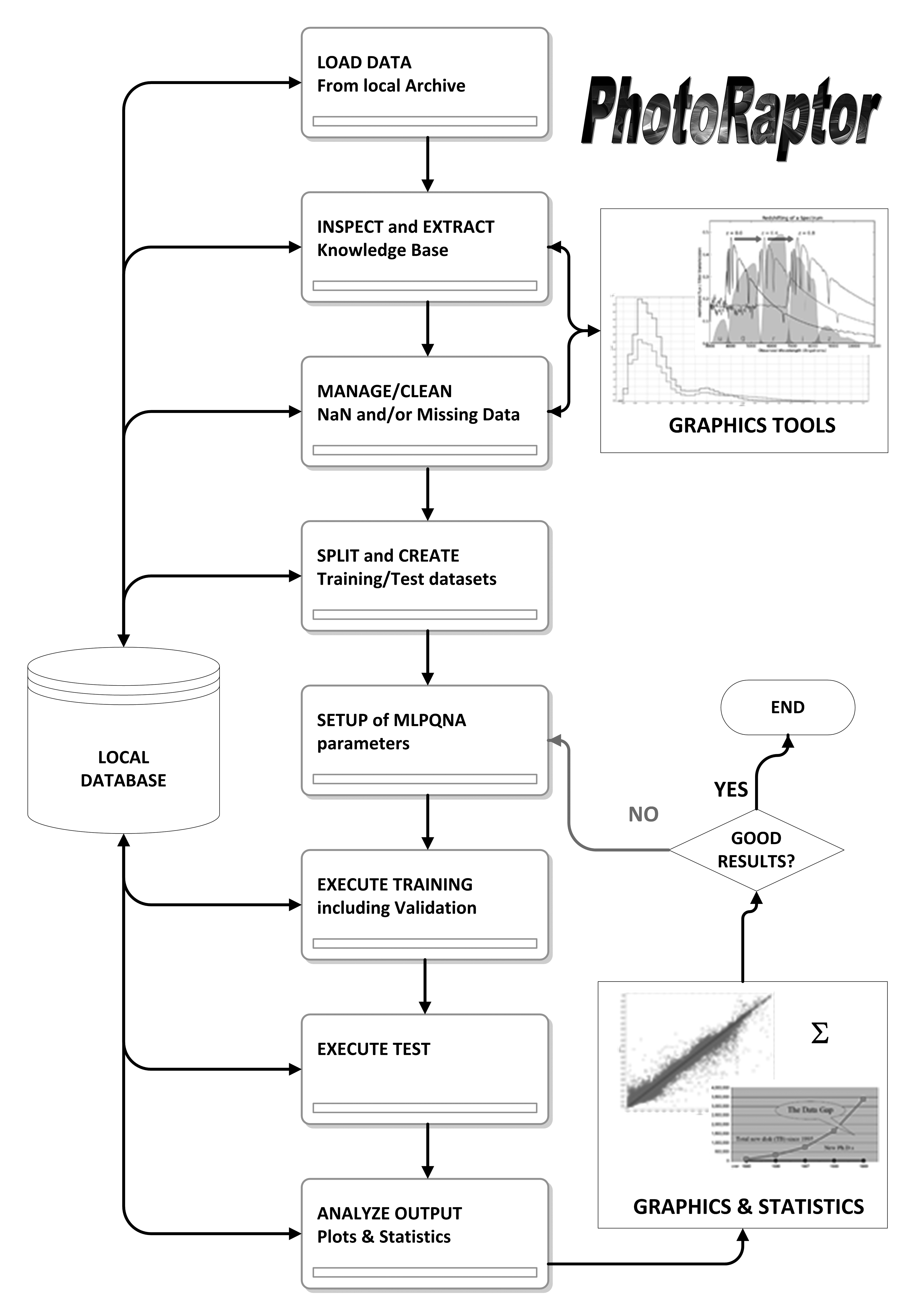}
\caption{The workflow of a generic experiment performed with PhotoRApToR.}
\label{fig:workflow}
\end{figure*}

Everyone who has used neural methods to produce photometric redshift evaluation knows that, in order to optimize the results in terms of
features, neural network architecture, evaluation of the internal and external errors, many experiments are needed. When coupled with the
needs of modern surveys, which require huge data sets to be processed, it clearly emerges the need for a user friendly, fast and scalable
application. This application needs to run client-side, since a great part of astronomical data is stored in private archives that are not
fully accessible on line, thus preventing the use of remote applications, such as those provided by the DAMEWARE tool (Brescia et al. 2014, \cite{brescia2014a}).
The code of the application was developed in Java language and runs on
top of a standard Java Virtual Machine, while the machine learning model was implemented in C++ language to increase the core execution
speed. Therefore different installation packages are provided to support the most common platforms.
Moreover, the application includes a wizard, which can easily introduce the user through the various functionalities offered by the tool.
The Fig.~\ref{fig:main} shows the main window of the program.
The main features of PhotoRApToR can be summarized as it follows:
\begin{itemize}
\item \textit{Data table manipulation}. It allows the user to navigate throughout his/her data sets and related \textit{metadata}, as well
as to prepare data tables to be submitted for experiments. It includes several options to perform the editing, ordering, splitting and
shuffling of table rows and columns. A special set of options is dedicated to the missing data retrieval and handling, for instance Not-a-Number
(NaN) or not calculated/observed parameters in some data samples;
\item \textit{Classification experiments}.
The user can perform general classification problems, i.e. automatic separation of an ensemble of data by assigning a common label to an
arbitrary number of their subsets, each of them grouped on the base of a hidden similarity.
The classification here is intended as \textit{supervised}, in the sense that there must be given a subsample of data for which the right
output label has been previously assigned, based on the \textit{a priori} knowledge about the treated problem. The application will learn on this
known sample to classify all new unknown instances of the problem;
\item \textit{Regression experiments}. The user can perform general regression problems, i.e. automatic learning to find out an embedded and
unknown analytical law governing an ensemble of problem data instances (patterns), by correlating the information carried by each element
(features or attributes) of the given patterns. Also the regression is here intended in a \textit{supervised} way, i.e. there must be given a
subsample of patterns for which the right output is \textit{a priori} known. After training on such KB, the program will be able to
apply the hidden law to any new pattern of the same problem in the proper way;
\item \textit{Photo-z estimation}. Within the \textit{supervised} regression functionality, the application offers a specialized toolset, specific
for photometric redshift estimation. After the training phase, the system will be able to predict the right photo-z value for any new sky object
belonging to the same type (in terms of photometric input features) of the Knowledge Base;
\item \textit{Data visualization}. The application includes some $2D$ and $3D$ graphics tools, for instance multiple histograms and multiple
$2D$/$3D$ scatter plots. Such tools are often required to visually inspect and explore data distributions and trends;
\item \textit{Data statistics}. For both classification and regression experiments a statistical report is provided about their output. In the
first case, the typical confusion matrix (Stehman 1997, \cite{stehman1997}) is given, including related statistical indicators such as classification efficiency,
completeness, purity and contamination for each of the classes defined by the specific problem. For what the regression is concerned, the application
offers a dedicated tool, able to provide several statistical relations between two arbitrary data vectors (usually two columns of a table), such as
average (bias), standard deviation ($\sigma$), Root Mean Square (RMS), Median Absolute Deviation (MAD) and the \textit{Normalized} MAD (NMAD,
Hoaglin et al. 1983, \cite{hoaglin1983}), the latter specific for the photo-z quality estimation, together with percentages of \textit{outliers} at the common threshold
$0.15$ and at different multiples of $\sigma$ (Brescia et al. 2014, \cite{brescia2014b}), (Ilbert et al. 2009, \cite{ilbert2009}).
\end{itemize}
In Fig.~\ref{fig:workflow} the layout of a general PhotoRApToR experiment workflow is shown. It is valid for either regression and classification cases.
%

\subsection{The Machine Learning model}
\label{mlpqna}

\begin{figure*}
\centering
\includegraphics[width=6cm]{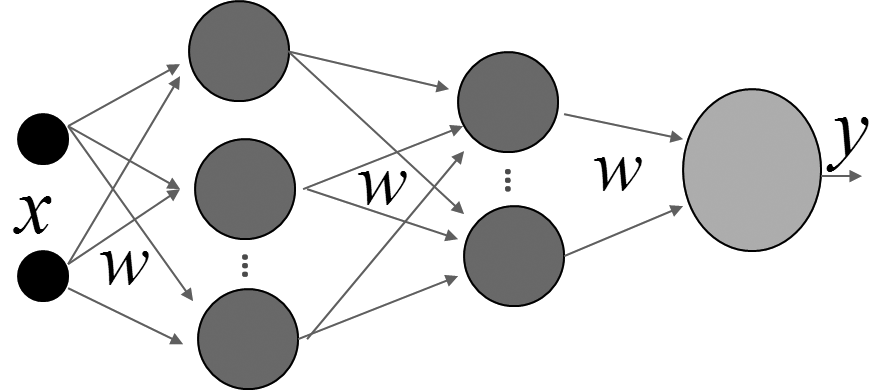}
\caption{The typical topology of a generic feed-forward neural network, in this case representing the architecture of MLPQNA.
In the simple example there are two hidden layers (the two blocks of dark gray circles) between the input (X) and output (Y) layers, corresponding to the architecture mostly used in the case of photo-z estimation. Arrows between layers indicate the connections (weights \textit{w}) among neurons. These weights are changed during the training iteration loop, according to the learning rule QNA.}
\label{fig:mlpqna}
\end{figure*}

The core of the PhotoRApToR application is its ML model, for instance the MLPQNA method. It is a Multi Layer
Perceptron (MLP; Rosenblatt 1961, \cite{rosenblatt1961}) neural network (Fig.~\ref{fig:mlpqna}), which is among
the most used feed-forward neural networks in a large variety of scientific and social contexts. The MLP is trained by a learning rule based on the Quasi Newton Algorithm (QNA).

The QNA is a variable metric method for finding local maxima and minima of functions (Davidon 1991, \cite{davidon1991}). The model
based on this learning rule and on the MLP network topology is then called MLPQNA.
QNA is based on Newton's method to find the stationary (i.e. the zero gradient) point of a function. In particular, the QNA is an optimization
of Newton's learning rule, because the implementation is based on a statistical approximation of the Hessian of the error function, obtained
through a cyclic gradient calculation.

In PhotoRApToR the Quasi Newton method was implemented by following the known L-BFGS algorithm (Limited memory - Broyden Fletcher Goldfarb Shanno;
Byrd 1994, \cite{byrd1994}), which was originally designed for problems with a very large number of features (hundreds to thousands), because in this case
it is worth having an increased iteration number due to the lower approximation precision because the overheads become much lower. This is
particularly useful in astrophysical data mining problems, where usually the parameter space is dimensionally huge and is often afflicted by a low
signal-to-noise ratio.

The analytical description of the method has been described in the contexts of both classification (Brescia et al. 2012, \cite{brescia2012b}) and regression
(Brescia et al. 2013, \cite{brescia2013} and Cavuoti et al. 2012, \cite{cavuoti2012}). In the present work, we focus the attention on its parameter setup and correct use within the presented framework.

\section{Photometric redshift estimation}
\label{photoz}
%
In practice, the problem of photo-z evaluation consists in finding the unknown function which maps the photometric set of features (magnitudes and/or
colors) into the spectroscopic redshift space. If a consistent fraction of the objects with spectroscopic redshifts is available, the problem can in
fact be approached as a data mining regression problem, where the a priori knowledge (i.e. the spectroscopic redshifts forming the KB), is used to
uncover the mapping function. This function can then be used to derive photo-z for objects without the spectroscopic counterpart information.
Without entering into much details, which can be found in the literature quoted below and in the references therein, we just outline that our method
has been successfully used in many experiments done on different KBs, often composed through accurate cross-matching among public surveys, such as SDSS
for galaxies (Brescia et al. 2014, \cite{brescia2014b}), UKIDSS, SDSS, GALEX and WISE for quasars (many of the following figures are referring to this experiment; Brescia et al. 2013, \cite{brescia2013}),
GOODS-North for the PHAT1 contest (Cavuoti et al. 2012, \cite{cavuoti2012}) and CLASH-VLT data for galaxies (Biviano et al. 2013, \cite{biviano2013}).
Other photo-z prediction experiments are in progress as preparatory work for the Euclid Mission (Laureijs et al. 2011, \cite{laureijs2011}) and the KiDS\footnote{http://www.astro-wise.org/projects/KIDS/} survey projects.
%
\subsection{User data handling}
\label{data}
%
The fundamental premise to use PhotoRaPToR is that the user must preliminarily know how to represent the data and, as trivial as it might seem,
it is worth to explicitly state that the user must: \textit{(i)} be conscious of the target of his experiment, such as for instance a regression
or classification; and \textit{(ii)} possess a deep knowledge of the used data. In what follows we shall call features the input parameters (i.e.,
for instance, fluxes, magnitudes or colors in the case of photo-z estimation).\\

\noindent \textit{Data Formats}\\

In order to reach an intelligible and homogeneous representation of data sets, it is mandatory to preliminarily take care of their internal format,
to transform the pattern features, and to force them to assume a uniform representation before submitting them to the training process. In this respect
real working cases might be quite different.
PhotoRApToR can ingest and/or produce data in any of the following supported formats:
\begin{itemize}
\item	FITS \cite{wells1981}: tabular/image;
\item	ASCII \cite{ansi1977}: ordinary text, i.e. space separated values;
\item	VOTable\footnote{http://www.ivoa.net/documents/VOTable/}: VO (Virtual Observatory) compliant XML-based documents;
\item	CSV \cite{repici2010}: Comma Separated Values;
\item   JPEG \cite{pennebaker1993}: Joint Photographic Expert Group, as image output type.
\end{itemize}

\noindent \textit{Missing Data}\\

Very frequently, data tables have empty entries (sparse matrix) or missing (lack of observed values for some features in some patterns).
Missing values (Marlin 2008, \cite{marlin2008}) are frequently (but not always) identified by special entries in the patterns, like Not-A-Number, out-of-range,
negative values in a numeric field normally accepting only positive entries etc.
Missing data is among the most frequent source of perturbation in the learning process, causing confusion in classification experiments or
mismatching in regression problems.
This is especially true for astronomy where inaccurate or missing data are not only frequent, but very often cannot be simply neglected since
they carry useful information. To be more specific, missing data in astronomical databases can be of two types:
\begin{itemize}
\item Type I: true missing data which were not collected. For instance a given region of the sky or a single object was not observed in a
given photometric band, thus leading to a missing information. These missing data may arise also from the simple fact that data, coming from
any source and related to a generic experiment, are in most case not expressly collected for data mining purposes and, when originally
gathered, some features were not considered relevant and thus left unchecked;
\item Type II: upper limits or non-detections (i.e. object too faint to be detected in a given band). In this case the missing datum conveys
very useful information which needs to be taken into account into the further analysis. It needs to be noticed, however that, often upper limits
are not measured in absence of a detection and therefore this makes these missing data undistinguishable from Type I.
\end{itemize}

In other words, missing data in a data set might arise from unknown reasons during data collecting process (Type I), but sometimes there are very
good reasons for their presence in the data since they result from a particular decision or as specific information about an instance for a subset
of patterns (Type II). This fact implies that special care needs to be put in the analysis of the possible presence (and related causes) of missing
values, together with the decision on how to submit these missing data to the ML method in order to take into account such special cases and prevent
wrong behaviors in the learning process.

Data entries affected by missing attributes, i.e. patterns having fake values for some features, may be used within the knowledge base used for a
photo-z experiment.
In particular they can be used to differentiate the data sets with an incremental quantity of affected patterns, useful to evaluate their noise
contribution to the performance of the photo-z estimation after training. Theoretically it has to be expected that a greater amount of missing data,
evenly distributed in both training and test sets, induces a greater deterioration in the quality of the results. This precious information may be
indeed used to assign different indices of quality to the produced photo-z catalogues.
The organization of data sets with different rates of missing data can be performed through PhotoRApToR by means of a series of options.\\

\begin{figure*}
\centering
\includegraphics[width=12cm]{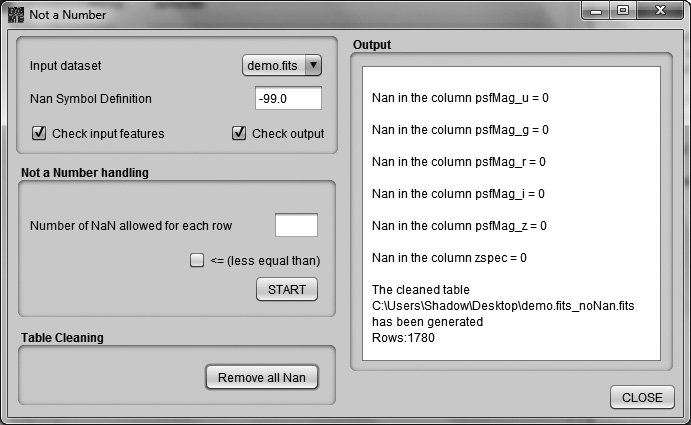}
\caption{Use of the NaN handling tool. After the definition of the NaN symbols, the user can generate a new dataset only with rows containing NaN
elements or another one cleaned by the NaN presence.}
\label{fig:NaN}
\end{figure*}

The Fig.~\ref{fig:NaN} shows the panel dedicated to define and quantify the presence of missing or bad data within the user tables.
The panel allows: (\textit{i}) to quantify the number of wrong values to be retained/removed in/from the data patterns; (\textit{ii}) to completely remove the data patterns affected by the presence of $NaN$ occurrences; (\textit{iii}) to assign arbitrary symbols to wrong or missing entries in the dataset (i.e. symbols like ``$-999$'', ``$NaN$'' or whatever).\\

\noindent \textit{Data Editing}\\

At the PhotoRApToR core is the MLPQNA neural model. In this respect, before launching any experiment, it may be necessary to manipulate data in order to
fulfill the requirements in terms of training and test patterns (data set rows) and features (data set columns) representation as well as contents:
\textit{(i)} both the training and test data files must contain the same number of input and target columns, and the columns must be in the same order;
\textit{(ii)} the target columns must always be the last columns of the data file;
\textit{(iii)} the input columns (features) must be limited to the physical parameters, without any other type of additional columns (like column
identifiers, object coordinates etc.);
\textit{(iv)} all input data must be numerical values (no categorical entries are allowed).

\begin{figure}
\centering
\includegraphics[width=15cm,angle=90]{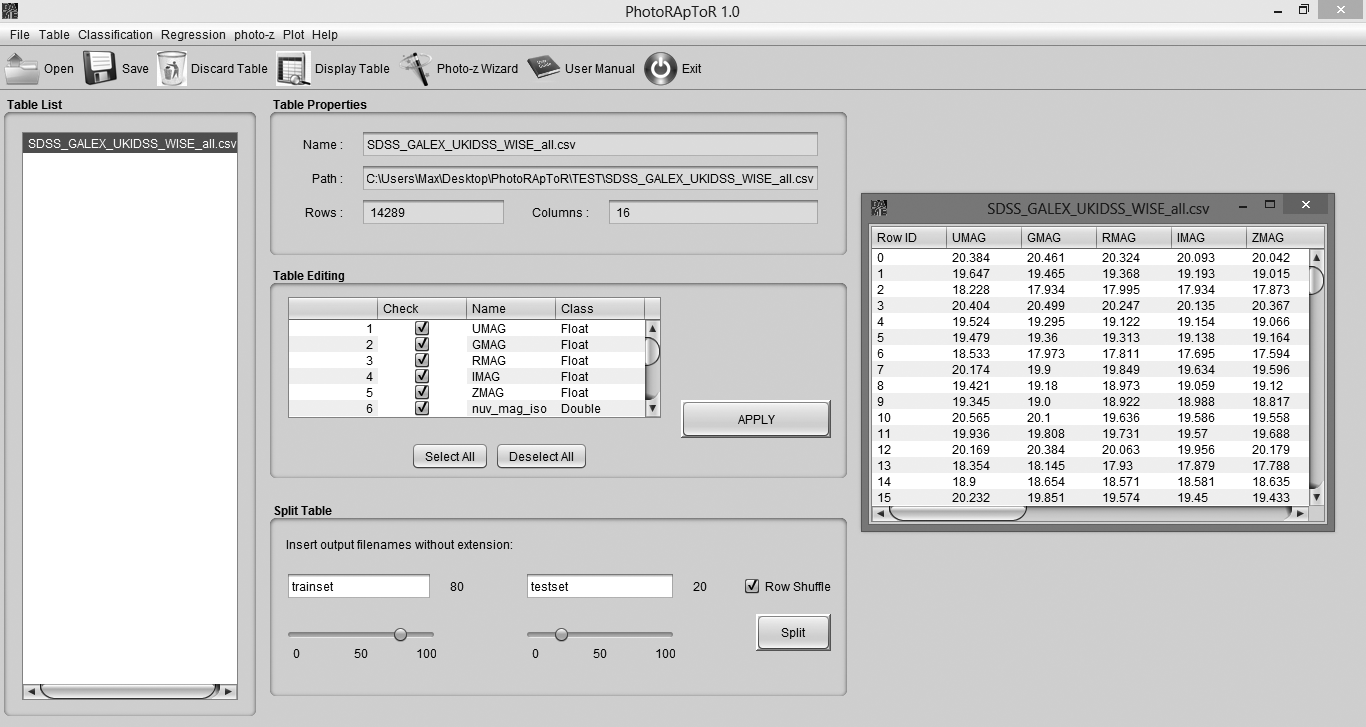}
\caption{The main panel showing details about the loaded data table and the editing options.}
\label{fig:mainedit}
\end{figure}

The application makes available a set of specific options to inspect and modify data file entries. Every time a new data table is loaded, a new window
shows the complete table properties (Fig.~\ref{fig:mainedit}), for instance: name, metadata, path and the number of columns and rows.

For a currently loaded table it is possible to select a subset of the needed columns.
After the selection, a table subset is created and, if the option \textit{Row Shuffle} is enabled, the subset rows are also randomly shuffled. The
random shuffling operation is useful to avoid systematic trends during the training phase and to ensure the homogeneity in the distribution of training
and test patterns.
This last property is, in fact, directly connected to the necessity to split the initial data into disjoint data sets,
to be used for the training and testing phases, respectively.
This is a simple action made possible by the \textit{Split} option. When the table is selected in the \textit{Table List}, the user must give two
different names for the split files (in this case \textit{train} and \textit{test}) and two different percentages of the original data set.
It is important to observe that, generally speaking, in machine learning supervised methods three different subsets for every experiment are
generally required from the available KB: \textit{(i)} the \textit{training set}, to train the method in order to acquire the hidden correlation
among the input features; \textit{(ii)} the \textit{validation set}, used to check and validate the training in particular against the loss of
generalization capabilities (a phenomenon also known as overfitting); and \textit{(iii)} the \textit{test set}, used to evaluate the overall
performances of the model (Brescia et al. 2013, \cite{brescia2013}).
In the version of the MLPQNA model implemented in the PhotoRApToR application, the validation is embedded into the training phase, by means of the
standard leave-one-out k-fold cross validation mechanism (Geisser 1975, \cite{geisser1975}).

Therefore, before any photo-z experiment, it is needed to split the data set in only two subsets, for instance, the training and test sets. There
is no any analytical rule to \textit{a priori} decide the percentages of the splitting operation. According to the direct experience, an empirical
rule of thumb suggests to use $80\%$ and $20\%$ for training and test sets, respectively (Kearns 1996, \cite{kearns1996}). But certainly it depends on the initial
amount of available KB. For example also $60\%$ vs $40\%$ and $70\%$ vs $30\%$ could be in principle used in case of large datasets (over ten
thousand patterns). The percentage depends also on the quality of the available KB. When both photometry and spectroscopy are particularly clean and
precise, with a high S/N, there could also be possible to obtain high performances by training just on half of the KB.

On the other hand, the more patterns are available for test, the more consistent will be the statistical evaluation of the experiment performances.\\

\noindent \textit{Data Plotting}\\

\begin{figure*}
\centering
\includegraphics[width=12cm]{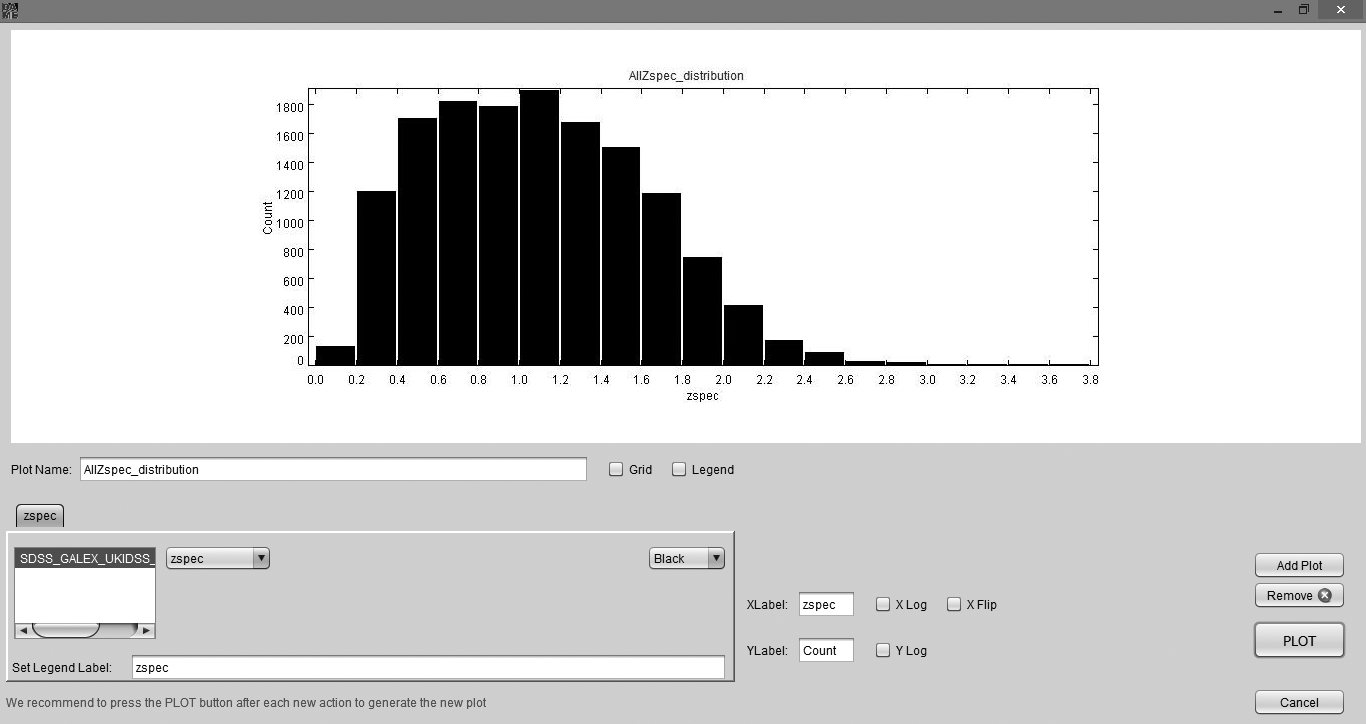}
\caption{An example of zspec distribution diagram, showing the options available within the histogram plotting panel.}
\label{fig:histosample}
\end{figure*}

\begin{figure*}
\centering
\includegraphics[width=12cm]{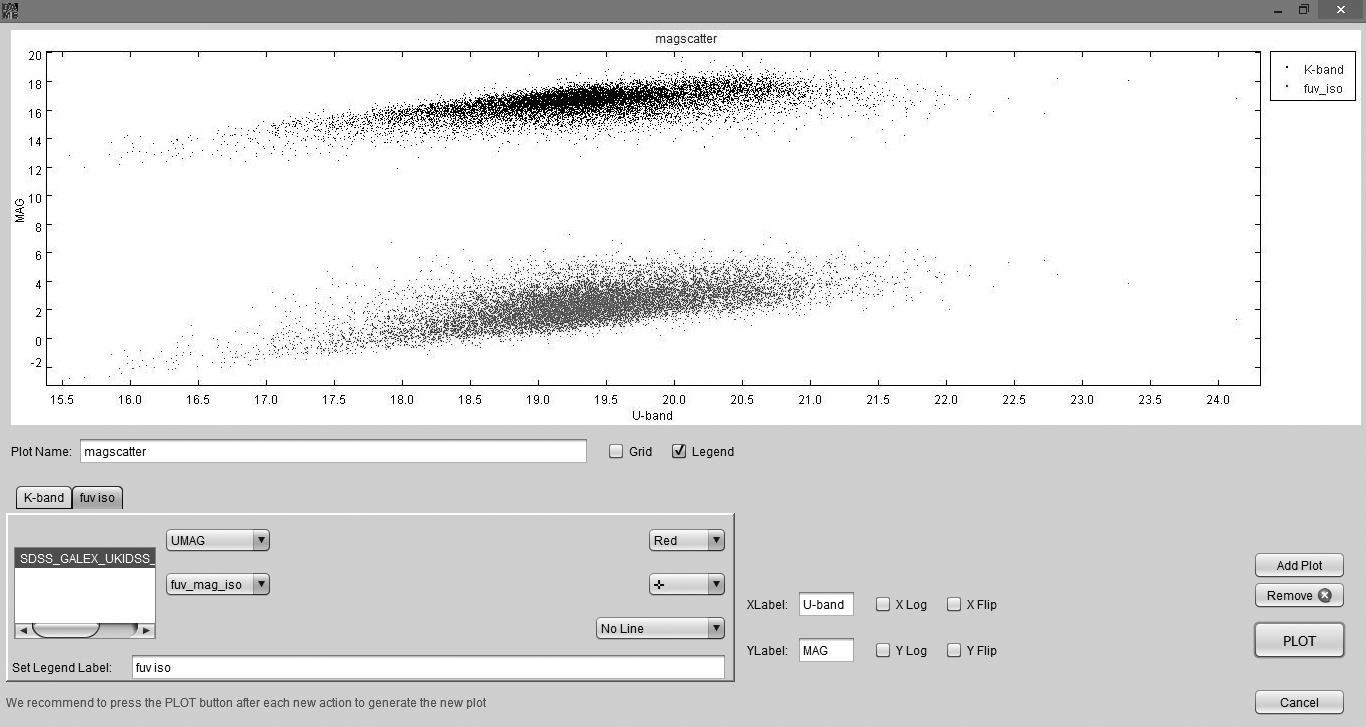}
\caption{An example of magnitude distributions, showing the options available within the 2D scatter plotting panel.}
\label{fig:scatter}
\end{figure*}

\begin{figure*}
\centering
\includegraphics[width=12cm]{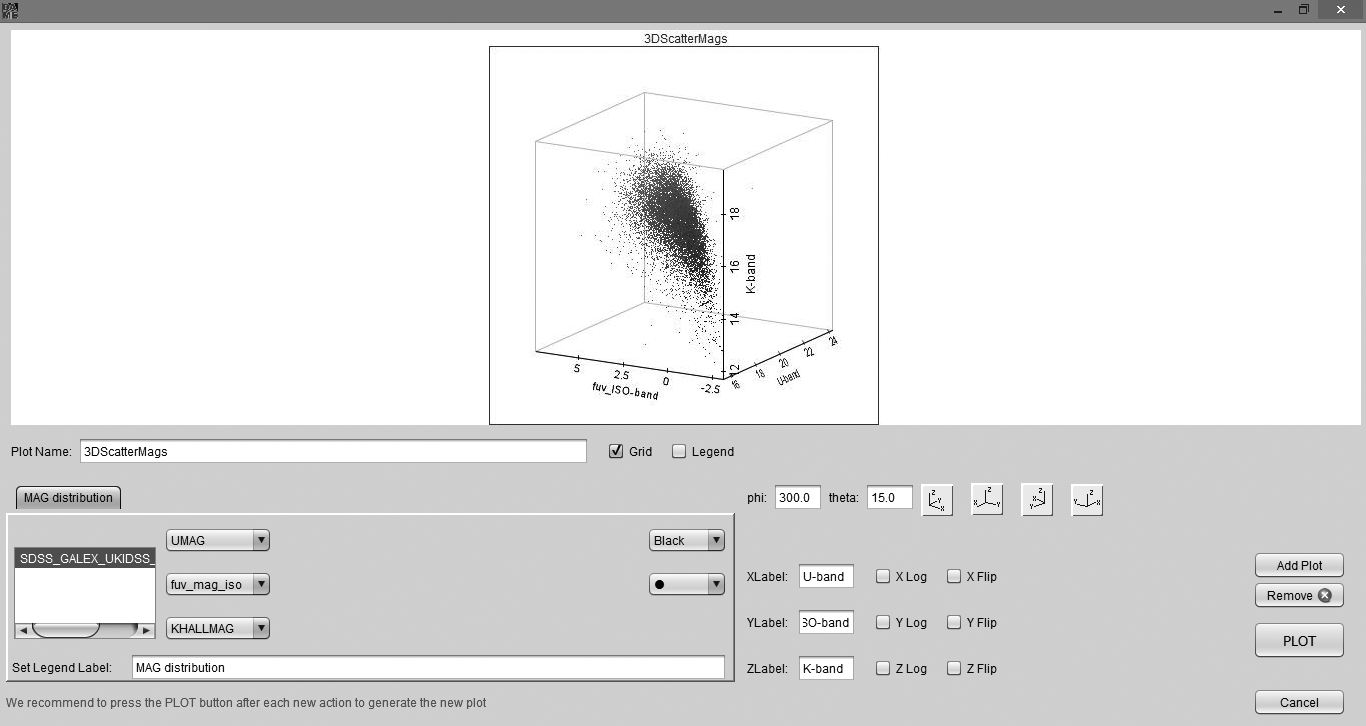}
\caption{An example of magnitude distributions, showing the options available within the 3D scatter plotting panel.}
\label{fig:3dscatter}
\end{figure*}

Within the PhotoRApToR application there are also instruments, based on STILTS toolset (Taylor 2006, \cite{taylor2006}), capable to generate different types of plots (some examples are shown in
Fig.~\ref{fig:histosample}, \ref{fig:scatter} and \ref{fig:3dscatter}). These options are particularly suited during the preparation phase of the
data for the experiments.

The graphical options selectable by user are:
\begin{itemize}
\item multi-column histograms;
\item multiple 2D and 3D scatter plots.
\end{itemize}

\noindent \textit{Data Feature Selection}\\

Learning by examples stands for a training scheme operating under supervision of an oracle capable to provide the correct, already known,
outcome for each of the training sample.
This outcome is properly a class or value of the examples and its representation depends on the available KB and on its intrinsic nature
even though in most cases it is based on a series of numerical attributes, related to the extracted KB, organized and submitted in an
homogeneous way.

Therefore, a fundamental step for any machine learning experiment is to decide which features to use as input attributes for the patterns
to be learned. In the specific case of photo-z estimation, for a given data sets, it is necessary to inspect and check which types of
fluxes (bands) and combinations (magnitudes, colors) is more effective.

In practice, the user must maximize the information carried by hidden correlations among different bands, magnitudes and zspec available.
In spite of what can be thought, not always the maximum number of available parameters should be suitable to train a machine learning model.
The experience demonstrates, in fact, that it is more the quality of data, than the quantity of features and patterns, the crucial key to obtain the
best prediction results (Brescia et al. 2013, \cite{brescia2013}). This phase is very time consuming and usually requires many tens or even hundreds of experiments.
Of course, the exact number of experiments depends on a variety of factors, among which, the number of photometric bands and magnitudes for
which a high quality of zspec entries is available in the KB; the photometric and spectroscopic quality of the data, the type of magnitudes
(i.e. aperture, total or isophotal magnitudes, etc.), the completeness of the spectroscopic coverage within the KB and the spectroscopic range.
In the authors experience, quite often, the optimal combination turned out to be the feature set obtained from the colors
plus one reference magnitude for each region of the electro-magnetic spectrum (broadly divided in UV, optical, Near Infrared,
Far Infrared, etc.) \cite{brescia2013}.
This can be understood by remembering that colors convey more information than the single related magnitudes, since from the basic equation defining
magnitudes it is easy to see that a magnitude difference corresponds to a flux ratio and hence in the derived colors an ordering relationship among features
is always implicitly assumed.

\subsection{Performing experiments}
\label{experiments}

After having prepared the KB, the user should have two subset tables ready to be submitted for a photo-z experiment. By looking at the
Fig.~\ref{fig:workflow} the experiment consists of a pre-determined sequence of steps, for instance:
\textit{(i)} Training and validation of the model network;
\textit{(ii)} blind Test of the trained model network;
\textit{(iii)} Run, i.e. the execution on new data samples of a well trained, validated and tested network.

We outline that for the first two steps, the basic rule is to use disjointed but homogeneous data subsets, because all empirical photo-z methods
in general may suffer to extrapolate outside the range of parameter distributions covered by the training. In other words, outside the limits
of magnitudes and spectroscopic redshift (zspec) imposed by the training set, these methods do not ensure optimal performances.
Therefore, in order to remain in a safe condition, the user must perform a selection of test data according to the training sample limits.

None of the objects included in the training sample should be included in the test sample and, moreover, only the data set used for the test
has to be used to generate performance statistics. In other words the test must be blind, i.e. containing only objects never submitted to the network before.

For what the training is concerned, this phase embeds two processing steps: the training of the MLPQNA model network and its validation.
It is in fact quite frequent for machine learning models to suffer of an \textit{overfitting} on training data, affecting and badly conditioning
the training performances. The problem arises from the paradigm of supervised machine learning itself. Any ML model is trained on a set of training
data in order to become able to predict new data points. Therefore its goal is not just to maximize its accuracy on training data, but mainly its
predictive accuracy on new data instances. Indeed, the more computationally stiff is the model during training, the higher would be the risk to fit
the noise and other peculiarities of the training sample in the new data \cite{dietterich1995}.
The technique implemented within PhotoRaPToR, i.e. the so called \textit{leave-one-out cross validation}, does not suffer of such drawback; it can
avoid overfitting on data and is able to improve the generalization performance of the ML model. In this way, validation can be implicitly performed
during training, by enabling at setup the standard leave-one-out k-fold cross validation mechanism \cite{geisser1975}.
The automatized process of the cross-validation consists in performing $k$ different training runs with the following procedure: (i) splitting of the
training set into $k$ random subsets, each one composed by the same percentage of the data set (depending on the $k$ choice); (ii) at each run the
remaining part of the data set is used for training and the excluded percentage for validation. While avoiding overfitting, the k-fold cross validation
leads to an increase of the execution time estimable around $k-1$ times the total number of runs.

\begin{figure*}
\centering
\includegraphics[width=15cm,angle=90]{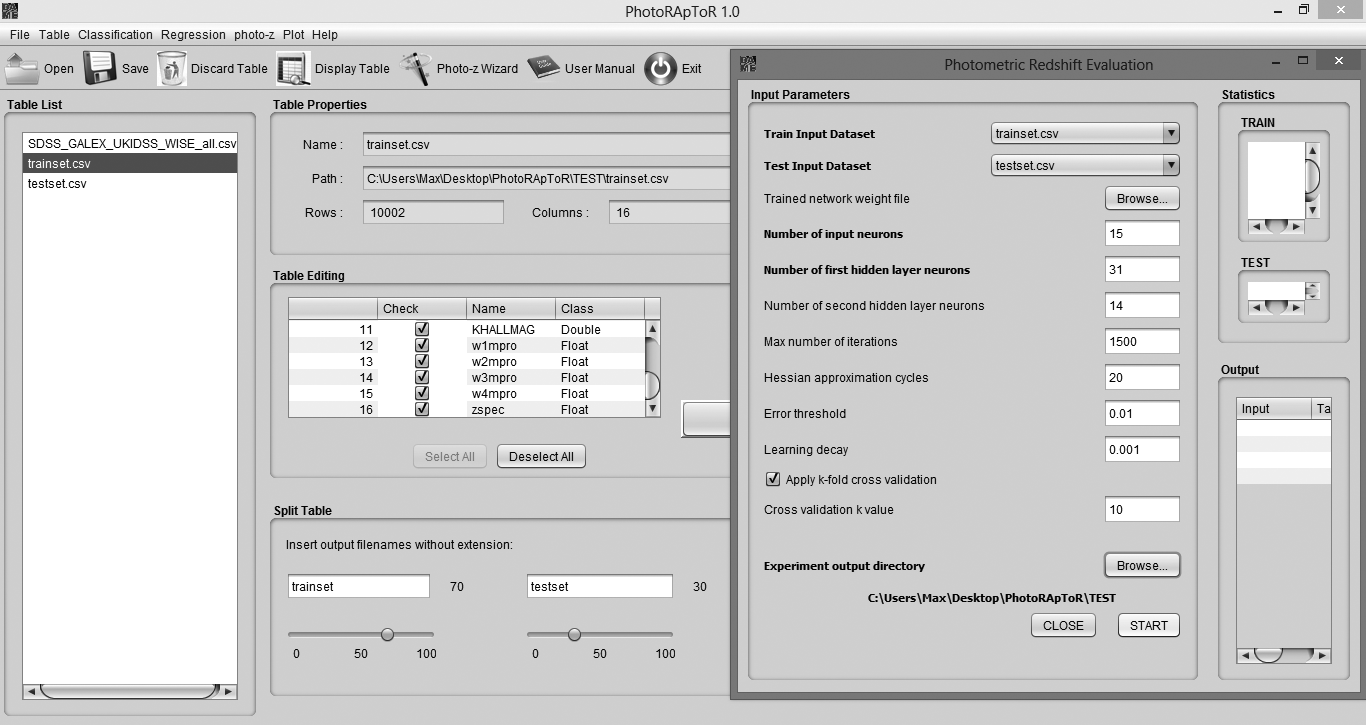}
\caption{An example of setup phase for a photo-z regression experiment.}
\label{fig:setup}
\end{figure*}

Concerning the photo-z experiment setup, special care must be paid to the setup of the training parameters, because all the other use cases, for
instance the Test and Run (i.e. the execution on new data), require only the specification of the proper input data set, and to recall the internal
model configuration as it was frozen at the end of training (Fig.~\ref{fig:setup}). We can group the MLPQNA model training parameters into three
subsets: \textit{network topology}, \textit{learning rule} setup and \textit{validation} setup.

\begin{itemize}
\item \textbf{Network topology}. It includes all parameters related to the MLP network architecture;
\begin{itemize}
\item \textit{Number of input neurons}. In terms of input data set it corresponds to the number of columns of the data table, (also named as input
features of the data sample, i.e. number of fluxes, magnitudes or colors composing the photometric information of each object in the data), except
for the target column (i.e. the spectroscopic redshift), which is related to the single output neuron of the regression network. More in general,
in the case of classification problems, the number of output neurons depends on the number of desired classes;
\item \textit{Number of neurons in the first hidden layer}. As a rule of thumb, it is common practice to set this number to $2N+1$, where N is the
number of input neurons. But it can be arbitrarily chosen by the user;
\item \textit{Number of neurons in the second hidden layer}. This is an optional parameter. Although not required in normal conditions, as stated
by the known universal approximation theorem \cite{cybenko1989}, some problems dealing with a parameter space of very high complexity, i.e. with
a large amount of distribution irregularities, are better treated by what was defined as \textit{deep} networks, i.e. networks with more than one
computational (hidden) layer \cite{bengio2007}. As a rule of thumb, it is reasonable to set this number to $N-1$, where $N$ is the number of input
neurons. But it is strongly suggested to use a number strictly lower than the dimension of the first hidden layer;
\item \textit{Number of neurons in the output layer}. This number is obviously forced to $1$ for regression problems, while in case of classification
this quantity depends on the number of classes as present within the treated problem;
\item \textit{Trained network weights}. This parameter is related to the matrix of weights (internal connections among neurons). A weight matrix exists
only after having performed one training session at least. Therefore, this parameter is left empty at the beginning of any experiment. But, for all
other use cases (Test or Run), it is required to load a previously trained network. However this parameter could also be used to perform further training
cycles for an already trained network (i.e. in case of an incremental learning experiment).
\end{itemize}
\item \textbf{Validation setup}: all parameters related to the optional training validation process;
\begin{itemize}
\item \textit{Cross validation k value}. When the cross validation is enabled, this value is related to the automatic procedure that splits in different
subsets the training data set, applying a k-step cycle in which the training error is evaluated and its performances are validated. Reasonable values
are between $5$ and $10$, depending on the amount of training data used. The k-fold cross validation intrinsically tries to avoid overfitting. Nonetheless, in rare cases (such as a wrong choice of the k parameter with respect to the train set dimension), a residual overfitting may occur. Therefore if the user wants to verify it, he/she should simply inspect the results, usually by comparing train with test performance. Whenever training accuracy is much better than test one, this is a typical clue of overfitting presence. Therefore, when cross validation with a proper k choice is enabled, by definition, it should avoid such events. The k parameter choice is not deterministic, but regulated by a rule of thumb, depending on the amount of training patterns. We remind also that this value strongly affects the overall computing time of the experiment.
\end{itemize}
\item \textbf{Learning rule setup}. It includes all parameters related to the QNA learning rule;
\begin{itemize}
\item \textit{Maximum number of iterations at each Hessian approximation cycle}.
The typical range for such value is $[1000, 10000]$, depending on the best compromise between the requested precision and the complexity of the problem. It can affect the computing time of the training;
\item \textit{ Number of Hessian approximation cycles}. Namely the number of approximation cycles searching for the best value close to the Hessian
of the error. If set to zero, the max number of iterations will be used for a single cycle. At each cycle the algorithm performs a series of iterations
along the direction of the minimum error gradient, trying to approximate the Hessian value. A reasonable range is $[20, 60]$, although also in this case
the exact value depends on the final precision required. If set to a high value, it is recommended to enable the cross validation option (see below),
to prevent overfitting occurrence;
\item \textit{Training error threshold}. This is one of the stopping criteria of the algorithm (alternative to the couple of the parameters
\textit{iterations} and \textit{cycles}). It is the training error threshold (a value of $0.001$ is typical for photo-z experiments).
\item \textit{Learning decay}. This value determines the analytical \textit{stiffness} of the approximation process. It affects the expression of
the weight updating law, by adding the term $decay*||network weights||^2$. Its range may vary from a minimum value of $0.0001$ (very low stiffness)
up to $1000.0$ (very high stiffness). Also in this case if a very low value is adopted, it is recommended to enable the cross validation option
(see below), to prevent overfitting occurrence.
\end{itemize}
\end{itemize}

The error calculated by the MLPQNA model during the training is evaluated for all the presented input patterns in terms of the difference between the
known target values and the calculated outputs of the model.
The error function in the regression case is based on the Least Mean Square (LSE) + Tychonov regularization \cite{groetsch1984}. This function is defined as follows:

\begin{eqnarray}
E = \frac{\sum^N_{i=1} (y_i-t_i)^2}{2} + \frac{decay*||W||^2}{2} \nonumber
\label{eqn:error}
\end{eqnarray}

\noindent where $N$ is the number of input patterns, $y$ and $t$ are the network output and the pattern target respectively, $decay$ is the decay
input parameter and $||W||$ the norm of the network weight matrix.\\

Regularization of the weight decay is the most important issue within the model mechanisms.
When the regularization factor is accurately chosen, then the generalization error of the trained neural network can be improved, and the training
can be accelerated. If the best regularization parameter $decay$ is unknown, it could be experimented by varying its value within the allowed range,
from a weak up to the strong regularization. In order to achieve the weight decay rule, we internally minimize a more complex merit function:

\begin{eqnarray}
f = E + \frac{decay*S}{2} \nonumber
\label{eqn:merit}
\end{eqnarray}

Here $E$ is the training error, $S$ is the sum of the squares of the network weights, and the
decay coefficient $decay$ controls the amount of smoothing applied to the network. Optimization is performed from the initial point and until the
successful stopping of the optimizer has been reached.\\
Searching for the best decay value is a typical trial-and-error procedure. It is usually performed by training the network with different values
of the parameter $decay$, from the lower value (no regularization) to the infinite value (strongest regularization).
By inspecting statistical results at each stage of the procedure  the overfitting tendency can be monitored by continuously changing the decay factor.
A zero decay usually corresponds to an overfitted network. Very large decay means instead an underfitted network. Between these extreme values there
is a range of networks which reproduce the dataset with different degrees of precision and smoothness.

After having successfully terminated a training session, the model will produce (among several output files) a final network weight matrix (file by
default called \textit{trainedWeights.txt}) and the network configuration setup (file by default called \textit{frozen\_train\_net.txt}), which can
be used during next experiment steps (Test and Run use cases), together with the respective input data sets.

\subsection{Inspection of results}
\label{results}

Interpolative methods, such as MLPQNA, have the advantage that the training set is made up of real objects.
In this sense, any empirical method intrinsically takes into account effects such as the filter band-pass and flux
calibrations, even though the difficulty in extrapolating to regions of the input parameter space not well sampled in the training data is one of the main
drawbacks \cite{collister2004}.

This is why a strong requirement of empirical methods is that the training set must be large enough to cover properly the parameter space in terms
of colors, magnitudes, object types and redshift. If this is true, then the calibrations and corresponding uncertainties are well known and only limited
extrapolations beyond the observed locus in color-magnitude space are required. Hence, under the conditions described above about the consistency of
the training set, a realistic way to measure photometric uncertainties is to compare the photometric redshifts estimation with spectroscopic measures
in the test samples.

All individual experiments should be evaluated in a consistent and objective manner through an homogeneous set of statistical indicators.
We remark that all statistical results reported throughout this paper are referred to the blind test data sets only. In fact, it is good practice to evaluate the results
on data (i.e. the test set) which have never been presented to the network during any of the training or validation phases. As easy to understand, the
combination of test and training data might introduce a straightforward systematic bias which could mask reality.

Within PhotoRApToR we use a specific algorithm to generate statistics.
For each experiment, given a list of $N$ blind test samples for $z_{spec}$ and $z_{phot}$, we define:
\begin{eqnarray}
 \Delta z &=& z_{spec} - z_{phot} \nonumber\\
 \Delta z_{norm} &=& \frac{z_{spec} - z_{phot}}{1 + z_{spec}} \nonumber
\end{eqnarray}

where $\Delta z_{norm}$ is the normalized $\Delta z$. By indicating with $x$ either $\Delta z$ or $\Delta z_{norm}$, we calculate the following
statistical indicators:
\begin{eqnarray}
 bias(x) &=& \frac{\sum^N_{i=1} x_i}{N} \nonumber \\
 \sigma (x) &=& \sqrt{\frac{\sum^N_{i=1} \left[x_i - \left(\frac{\sum^N_{i=1} x_i}{N}\right)\right]^2}{N}} \nonumber \\
 RMS(x) &=& \sqrt{\frac{\sum^N_{i=1} x_i^2}{N}} \nonumber \\
 MAD (x) &=& Median (\mid x \mid) \nonumber \\
 NMAD (x) &=& 1.4826 \times Median (\mid x \mid) \nonumber
\end{eqnarray}

There is also a relation between the Root Mean Square (RMS) and the Standard Deviation $\sigma$: $RMS = \sqrt{mean^2 + \sigma^2}$, but $\sigma^2$ is
the \textit{variance}, so we have $RMS = \sqrt{mean^2 + variance}$. Therefore, for a direct comparison of results, in terms of distance of
$m\sigma$ ($m = 1, 2,...$) from the distribution of $\Delta z$, it is much more precise to use the Standard Deviation as main indicator, rather than
the simple RMS.

There is often a confusion about the relation between photometric and spectroscopic redshifts used to apply the statistical indicators. For instance,
the performance could be very different if the simple $\Delta z$ is used instead of the $\Delta z_{norm}$. The idea is that the $\Delta z$ cannot
represent the best statistical indicator in the specific case of photometric redshift prediction.

The velocity dispersion error, intrinsically present within the photometric estimation, is not uniform over
a wide range of spectroscopic redshift and therefore the related statistics is not able to give a consistent estimation. On the contrary, the
normalized term $\Delta z_{norm}$ introduces a more uniform information, correlating in a more correct way the variation of photometric estimation,
and thus permitting a more consistent statistical evaluation at all ranges of spectroscopic redshift.

For what the analysis of the catastrophic outliers is concerned, according to \cite{mobasher2007}, the parameter
$D_{95} \equiv \Delta_{95}/\left(1+z_{phot}\right)$ enables the identification of outliers in photometric redshifts derived through SED fitting methods
(usually evaluated through numerical simulations based on mock catalogues).
In fact, in the hypothesis that the redshift error $\Delta z_{norm}$ is Gaussian, the catastrophic redshift error limit would be constrained by the width of the redshift probability distribution, corresponding to the $95\%$ confidence interval, i.e. with $\Delta_{95} = 2\sigma \left( \Delta z_{norm} \right)$.
In our case, however, photo-z are empirical, i.e. not based on any specific fitting model and it is preferable to use the standard deviation value
$\sigma \left( \Delta z_{norm} \right) $  derived from the photometric cross matched samples, although it could overestimate the theoretical Gaussian
$\sigma$, due to the residual spectroscopic uncertainty as well as to the method training error. Therefore, we consider as catastrophic outliers the
objects with $\left| \Delta z_{norm} \right| > 2 \sigma \left( \Delta z_{norm} \right)$. This although it is common practice to indicate as outliers
all objects with $\left| \Delta z_{norm} \right| > 0.15$, (thus included in the provided statistics).

It is also important to notice that for empirical methods it is useful to analyze the correlation between the
$NMAD\left( \Delta z_{norm} \right) = 1.48 \times median \left( \left| \Delta z_{norm} \right| \right)$ and the standard deviation
$\sigma_{clean}(\Delta z_{norm})$ calculated on the data sample for which $\left| \Delta z_{norm} \right| \leq 2 \sigma \left( \Delta z_{norm} \right)$.
In fact, the quantity $NMAD$ is smaller than the value of the $\sigma_{clean}$. In such condition we can assert that the pseudo-gaussian distribution of
$\left( \Delta z_{norm} \right)$ is mostly influenced by the presence of catastrophic outliers.

\begin{figure*}
\centering
\includegraphics[width=15cm,angle=90]{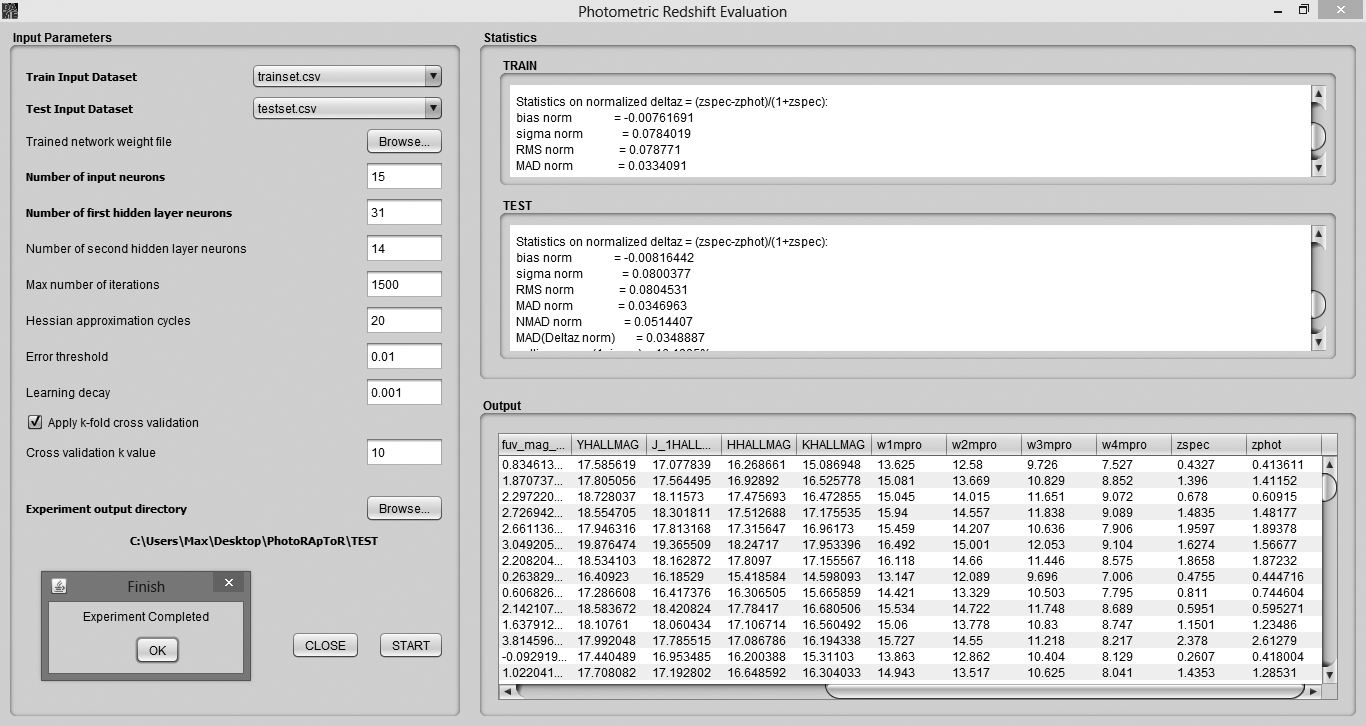}
\caption{The statistics produced at the end of a photo-z regression experiment. The training and test results are also automatically stored in the
files \textit{train\_statistics.txt} and \textit{test\_statistics.txt}, respectively.}
\label{fig:stats}
\end{figure*}

\begin{figure*}
\centering
\includegraphics[width=12cm]{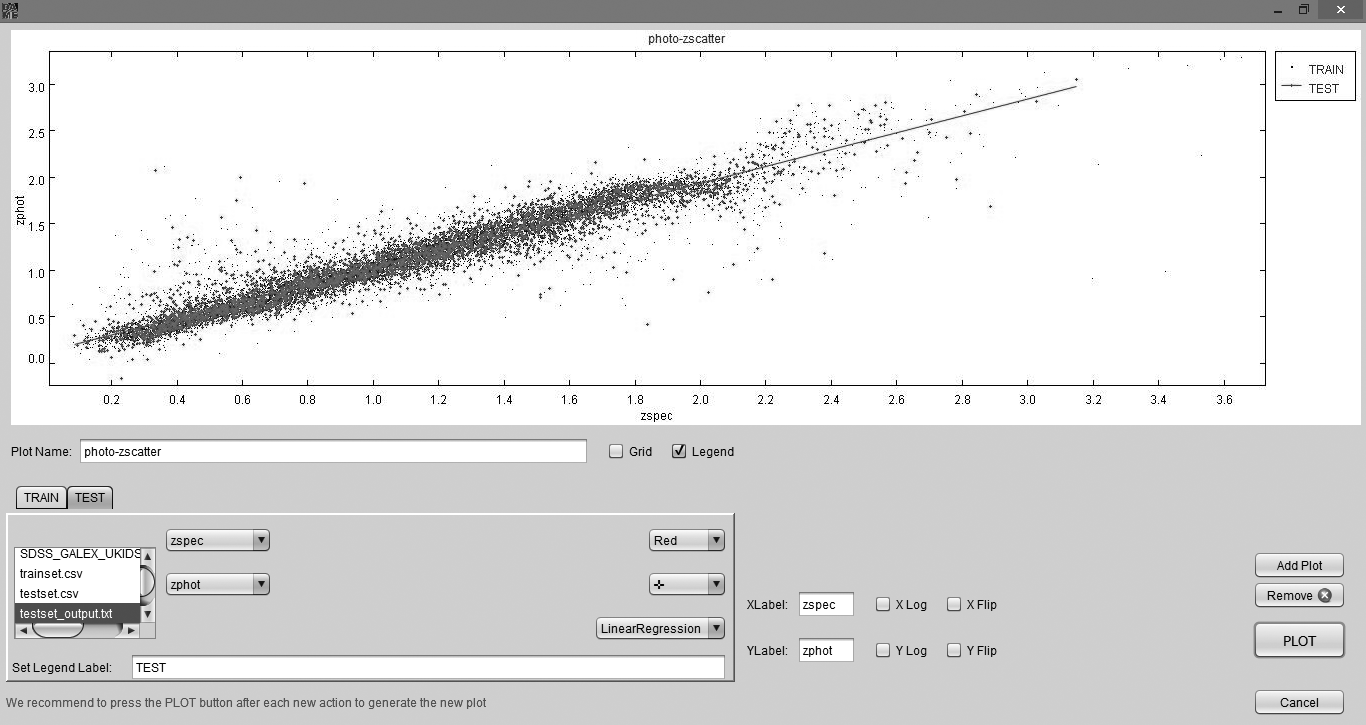}
\caption{The photo-z vs zspec plot as produced after a photo-z regression experiment. In this example the diagram shows both training (black dots)
and test (gray crosses) objects, although the blind test objects are the most relevant to evaluate the prediction performances.}
\label{fig:zphot}
\end{figure*}

All the described statistical indicators are provided by PhotoRaPToR as the output of every photo-z estimation test and are stored in specific files
(by default named as \textit{test\_statistics.txt}). For completeness we also provide a similar statistics file as the output of any training session
(Fig.~\ref{fig:stats}).
But its use is only to allow a quick comparison between training and test, just in order to verify the absence of any overfitting occurrence.\\
Besides the statistics files, PhotoRApToR makes also available some graphical tools, useful to perform a visual inspection of photo-z experiments.
In particular a $2D$ scatter plot to show the trend of photo-z vs zspec (Fig.~\ref{fig:zphot}), as well as a set of histograms useful to graphically
evaluate the distributions of quantities $\Delta z$ and $\Delta z_{norm}$.\\

\section{Other functionalities}
\label{others}

To complete the description of the resources made available by PhotoRApToR, we wish to stress that besides photometric redshift estimation (to be
intended as a specific type of regression experiment), the user has the possibility to perform generic regression as well as multi-class classification
experiments.

For a generic regression problem, all the above functionalities described in the case of the photo-z, remain still valid, with the only straightforward
exception for the statistics produced, which is generated for generic quantities formulated below.

\begin{eqnarray}
 \Delta out &=& target - output \nonumber\\
 \Delta out_{norm} &=& \frac{target - output}{1 + target} \nonumber
\end{eqnarray}

Also in the case of the multi-class classification, the above considerations and options remain still valid with only some differences, described in what follows.

\begin{figure*}
\centering
\includegraphics[width=15cm,angle=90]{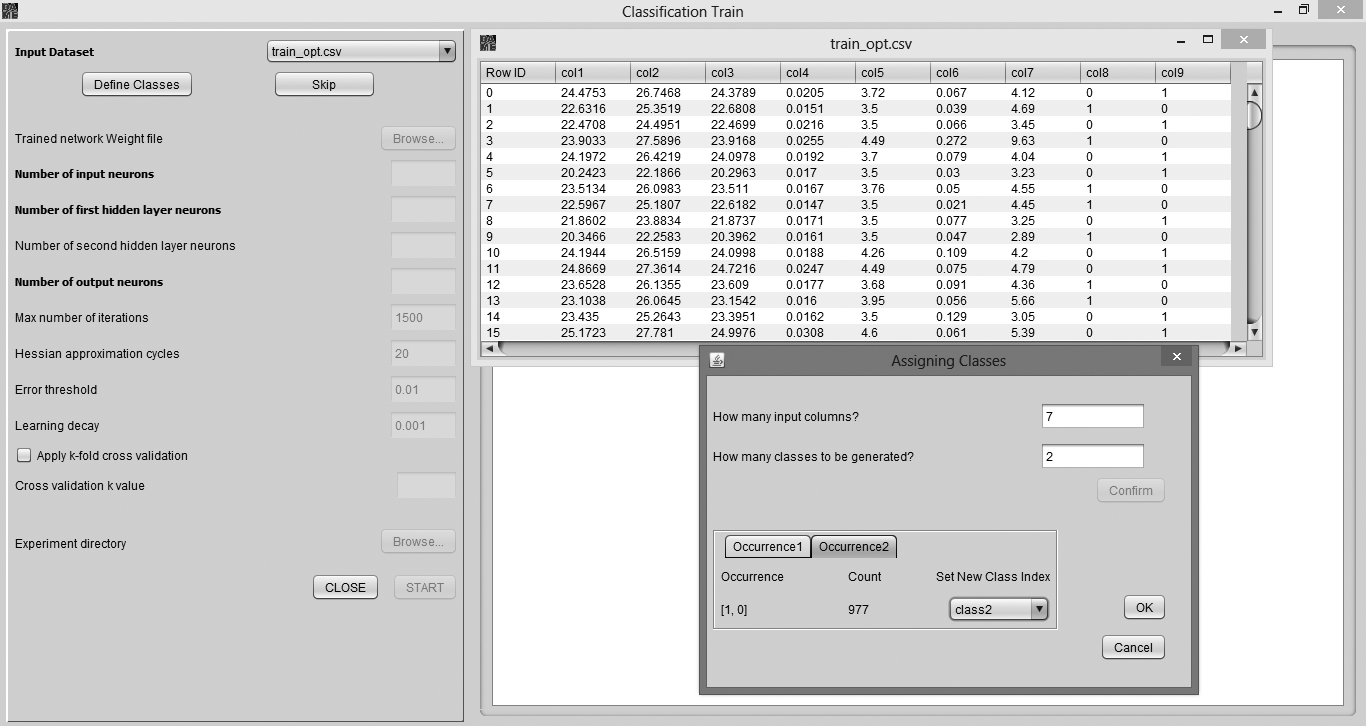}
\caption{The setup panel of a multi-class classification experiment. It is also possible to assign arbitrary class labels to all output instances in the
training and test sets (see subpanel \textit{Assigning Classes}).}
\label{fig:clsetup}
\end{figure*}

During the training setup (Fig.~\ref{fig:clsetup}), there are two specific options, not foreseen for regression problems:

\begin{itemize}
\item \textit{Output neurons}. The number of neurons of the output layer (which is forced to be $1$ in the regression experiments), in this case
corresponds to the number of different classes present in the training sample. It is required that the class identifiers should have a binary format
label. For instance, in a three-class problem, the target classes are represented in three columns labeled respectively, as $(1 0 0)$, $(0 1 0)$ and $(0 0 1)$;
\item \textit{Cross entropy}: this optional parameter, if enabled, replaces the standard training error evaluation (for instance the MSE between output
and target values). Its meaning is discussed below.
\end{itemize}

The Cross Entropy (CE) error function was introduced to address classification problem evaluation in a consistent statistical fashion \cite{rubin2004}.
The CE method consists of two phases:
\textit{(i)} it generates a random data sample (trajectories, vectors, etc.) according to a specified mechanism; %
\textit{(ii)} it updates the parameters of the random mechanism based on the data to produce a \textit{better} sample in the next iteration.

In practice, a data model is created based on the training set, and its CE is measured on a test set to assess how accurately the model is predicting
the test data. The method compares indeed two probability distributions, $p$ the true distribution of data in the data set, and $q$ which is the distribution
of data as predicted by the model. Since the true distribution is unknown, the CE cannot be directly calculated, while an estimate of CE is obtained using the
following expression:

$$H\left(T,q \right)=- \sum_{i=1}^{N} \frac{1}{N} log_2 q\left(x_i \right)$$

\noindent where $T$ is the chosen training set, corresponding to the true distribution $p$, $N$ is the number of objects in the test set, and
$q\left( x \right)$ is the probability of the event $x$ as estimated from the training set.

\begin{figure*}
\centering
\includegraphics[width=15cm,angle=90]{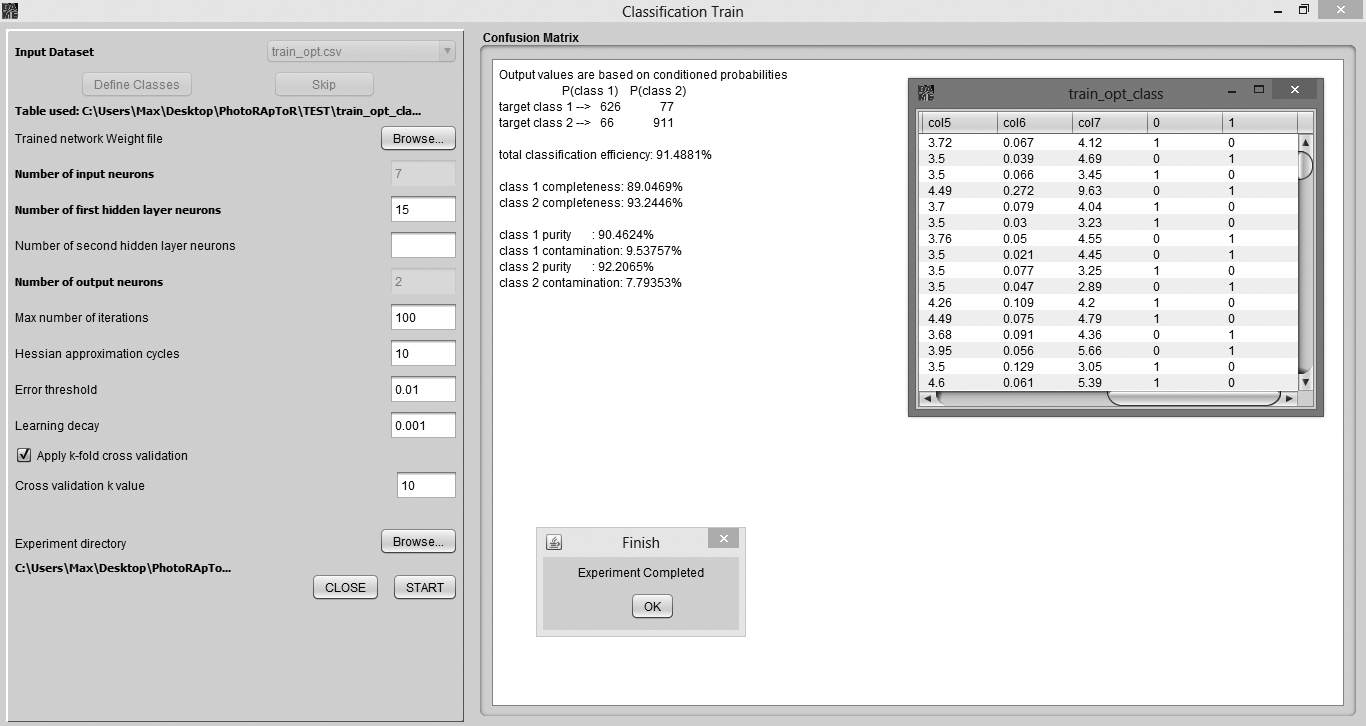}
\caption{The statistics produced at the end of a 2-class classification experiment.}
\label{fig:cloutput}
\end{figure*}

Another difference with respect to regression experiments is of course the statistics produced to evaluate the results outcoming from a classification experiment.
In this case, at the base of the statistical indicators adopted, there is the commonly known \textit{confusion matrix}, which can be used to easily visualize the
classification performance \cite{provost1998}: each column of the matrix represents the instances in a predicted class, while each row represents the instances
in the real class (Fig.~\ref{fig:cloutput}). One benefit of a confusion matrix is the simple way in which it allows to see whether the system is mixing different
classes or not.

More specifically, for a generic two-class confusion matrix,

\begin{eqnarray}
  \begin{array}{c|ccc}
            &            & $OUTPUT$\\ \hline
            &    -       &$Class A$      & $Class B$ \\
    $TARGET$  &$Class A$    & N_{AA}              & N_{AB} \\
            &$Class B$    & N_{BA}              & N_{BB} \\
  \end{array} \nonumber
  \end{eqnarray}

\noindent we then use its entries to define the following statistical quantities:

\begin{itemize}
\item \underline{total efficiency}: $te$. Defined as the ratio between the number of correctly classified objects and the total number of objects in the data set.
In our confusion matrix example it would be:
    \begin{eqnarray}
    te=\frac{N_{AA} + N_{BB}}{N_{AA} + N_{AB} + N_{BA} + N_{BB}} \nonumber
    \end{eqnarray}
\item \underline{purity of a class}: $pcN$. Defined as the ratio between the number of correctly classified objects of a class and the number of objects classified
in that class. In our confusion matrix example it would be:
    \begin{eqnarray}
    pcA=\frac{N_{AA}}{N_{AA}+N_{BA}} \nonumber
    \end{eqnarray}
    \begin{eqnarray}
    pcB=\frac{N_{BB}}{N_{AB}+N_{BB}} \nonumber
    \end{eqnarray}
\item \underline{completeness of a class}: $cmpN$. Defined as the ratio between the number of correctly classified objects in that class and the total number of
objects of that class in the data set. In our confusion matrix example it would be:
    \begin{eqnarray}
    cmpA=\frac{N_{AA}}{N_{AA}+N_{AB}} \nonumber
    \end{eqnarray}
    \begin{eqnarray}
    cmpB=\frac{N_{BB}}{N_{BA}+N_{BB}} \nonumber
    \end{eqnarray}
\item \underline{contamination of a class}: $cntN$. It is the dual of the purity, namely it is the ratio between the misclassified objects in a class and the number
of objects classified in that class, in our confusion matrix example will be:
    \begin{eqnarray}
    cntA=1-pcA=\frac{N_{BA}}{N_{AA}+N_{BA}} \nonumber
    \end{eqnarray}
    \begin{eqnarray}
    cntB=1-pcB=\frac{N_{AB}}{N_{AB}+N_{BB}} \nonumber
    \end{eqnarray}
\end{itemize}

All these statistical indicators are packed in an output file, produced at the end of the test phase of any classification experiment.

The MLPQNA machine learning method, embedded into PhotoRaptor, has been already tested in several classification cases. In Brescia et al. 2012, \cite{brescia2012b}, we compared the performances of MLPQNA with other machine learning based classifiers and traditional techniques as well, in terms of accuracy of identifying candidate globular clusters in the NGC 1399 HST single-band data. In Cavuoti et al. 2014, \cite{cavuoti2014}, we compared MLPQNA with standard MLP and Support Vector Machine to photometrically classify AGNs in the SDSS DR4 archive. Finally, we recently have exploited the MLPQNA to perform classification experiments within SDSS DR10 archive, aimed at photometrically identifying quasars from the whole sample including also galaxies and stars, as well as to verify the possibility to disentangle normal galaxies from objects with a peculiar spectrum, (Brescia et al. 2015, \cite{brescia2015}).

\section{Comparison with public machine learning tools}
\label{comparison}

We performed a simple comparison between PhotoRApToR and an alternative machine learning tool publicly available: the scikit-learn toolset \cite{pedregosa2011}.
The comparison is based on the photo-z estimation by means of a supervised non-linear regression experiment, by directly comparing the statistical performances between the MLPQNA model provided through PhotoRApToR and the widely known ensemble method based on Random Forest \cite{breiman2001}, which uses a random subset of candidate data features to build an ensemble of decision trees.

The data set used for the experiment was obtained by merging the photometry from four different surveys (UKIDSS, SDSS, GALEX and WISE), including derived colors and reference magnitudes for each band as internal features, thus covering a wide range of wavelengths from the UV to the mid-infrared. While the spectroscopic redshifts, (i.e. the zspec target values) were derived from selected quasars of the SDSS-DR7 database. The complete KB consisted of $\sim1.4\times10^4$ objects, from which the 60\% used as training set and the residual 40\% as blind test set (see Brescia et al. 2013, \cite{brescia2013}, for more details). We remark also that in that case, our MLPQNA has been directly compared with other several photo-z estimation methods (see references therein), achieving best results.

\begin{figure*}
\centering
\includegraphics[width=12cm]{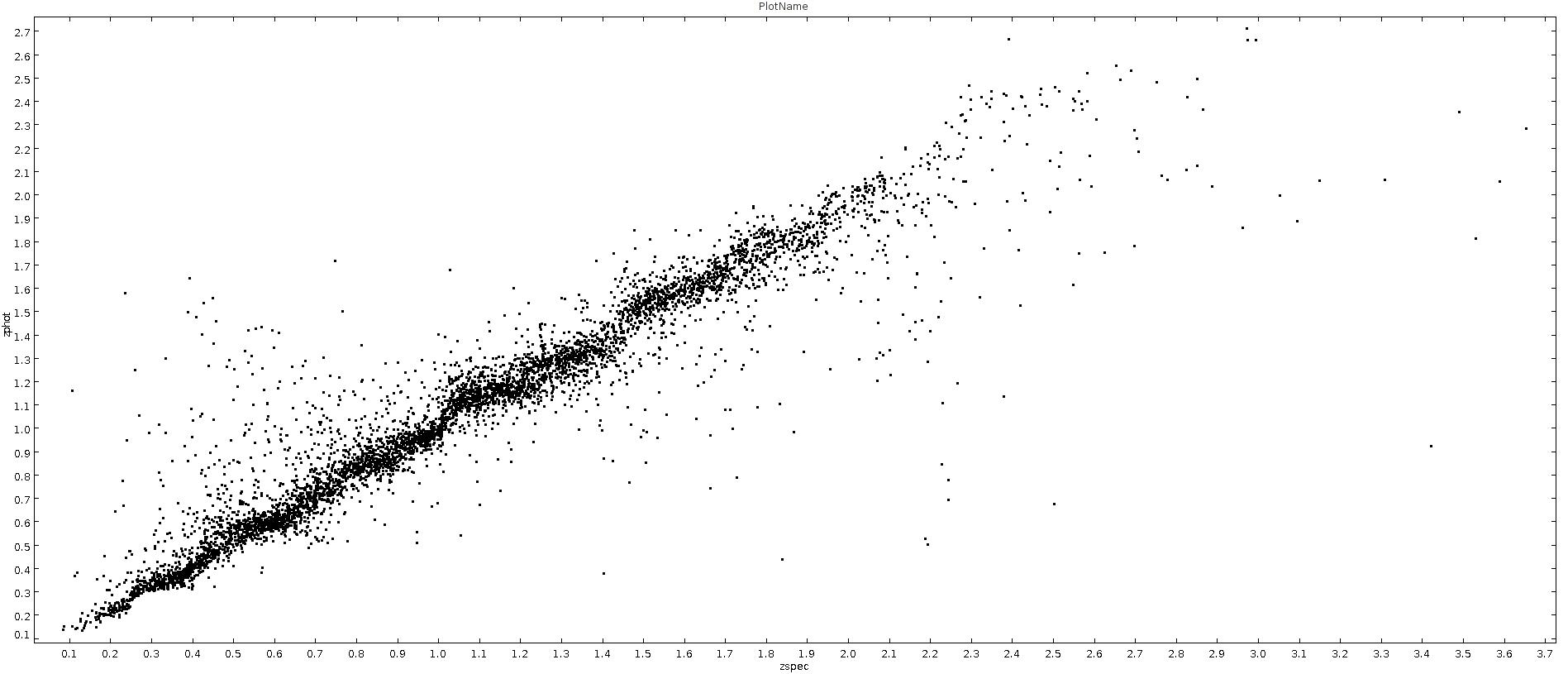}
\includegraphics[width=12cm]{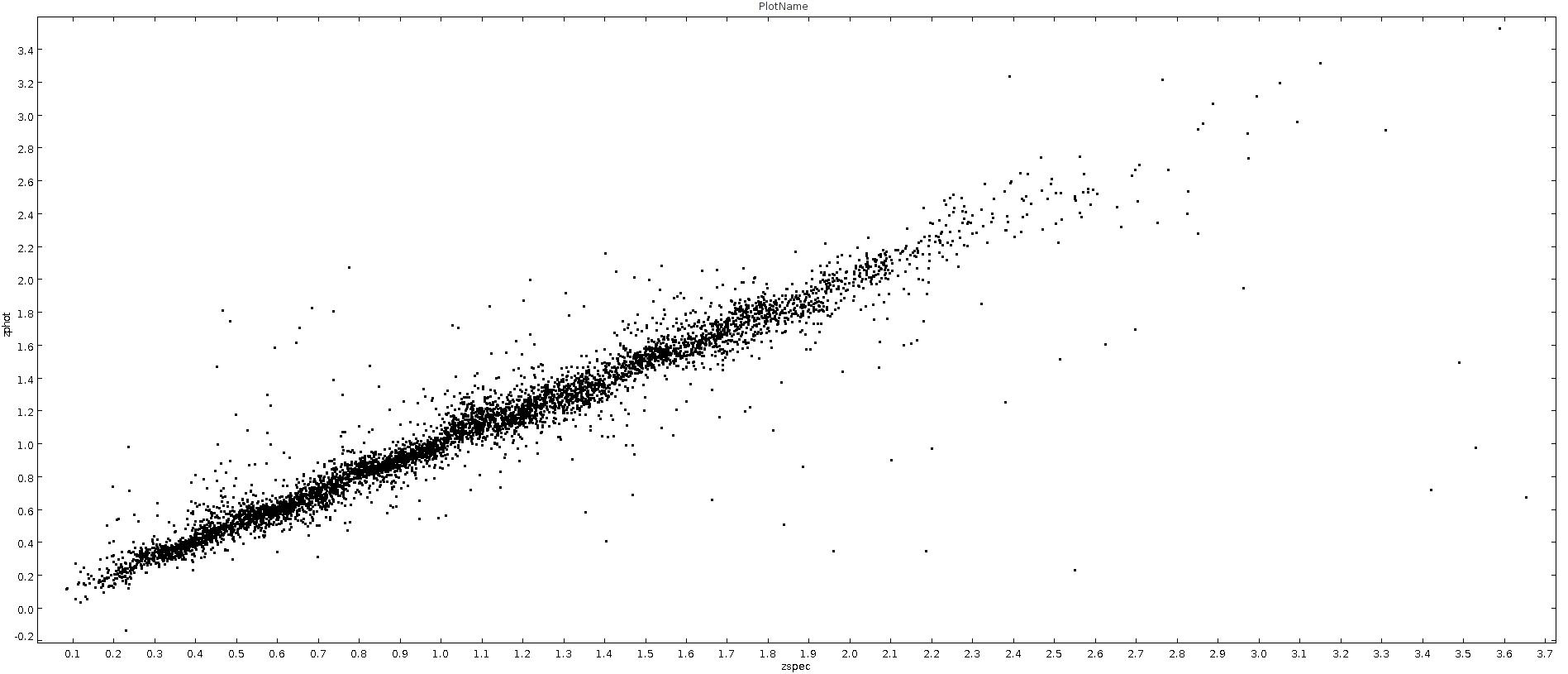}
\caption{The photo-z vs zspec scatter plots as produced after the photo-z estimation experiment. The upper plot refers to the Random Forest model while the lower one is related to the MLPQNA model results. Both diagrams show the distributions of the $\sim5.7\times10^3$ objects composing the blind test set.}
\label{fig:zphotcomp}
\end{figure*}

After having trained the two ML models with the same training set, their photo-z estimation results have been compared in terms of statistics and residual analysis (outlier percentages). The results are shown in Fig.~\ref{fig:zphotcomp} and reported in Tab.~\ref{qso:tab:comparison}.
From the comparison, it results apparent that MLPQNA performs better than Random Forest, especially in the high-redshift zone (i.e. at $zspec>2.0$), showing a more robust prediction capability also in the sparsely populated regions of the parameter space.

In addition, unlike the PhotoRApToR resource, in order to setup and run the Random Forest model provided by the scikit-learn package, as well as to prepare and execute the experiments, some manipulations of the source code have been necessary. The reason is that the scikit-learn package is provided as a library to be imported in a user-defined script code, which implies a certain knowledge of the Python programming language.

\begin{table*}
\begin{center}
\caption{Comparison of the performances among the different tools. MLPQNA is the ML engine of our application, based on a four-layers neural network, while Random Forest is the ML model provided by the scikit-learn public resource. Both methods were trained on the multi-survey mixed (colors + reference magnitudes) dataset, obtained by cross-matching photometry of UKIDSS, SDSS, GALEX and WISE surveys. The reported statistics is related to the photo-z estimation on the blind test set of about $\sim5.7\times10^3$ QSO objects. For the definition of the parameters and for discussion see text.}\label{qso:tab:comparison}
\begin{tabular}{|c|c|c|c|c|c|} \hline
Photo-z Estimation Statistics &        &          & $\Delta z_{norm}$ \\ \hline
Model                         & $BIAS$ & $\sigma$ & $MAD$ & $RMS$ & $NMAD$ \\ \hline
PhotoRApToR (MLPQNA)          & 0.004  & 0.069 & 0.020 & 0.069 & 0.029 \\ \hline
Scikit (Random Forest)        & 0.009  & 0.083 & 0.021 & 0.084 & 0.031 \\ \hline \hline
Outlier percentages $[\%]$    &        &          & $|\Delta z_{norm}|$ \\ \hline
Model                         & $> 0.15$ & $> 1\sigma$ & $> 2\sigma$ & $> 3\sigma$ & $> 4\sigma$ \\ \hline
PhotoRApToR (MLPQNA)          & 2.43     &  9.39       & 2.89        & 1.40        & 0.91 \\ \hline
Scikit (Random Forest)        & 5.27     & 11.03       & 4.48        & 2.31        & 1.34 \\ \hline
\end{tabular}
\end{center}
\end{table*}

Although we reported a use case example where PhotoRaptor has been tested on a dataset of about $10^4$ samples, we want to emphasize that the reliability of our resource has been already verified for data sets up to $\sim10^6$ samples. However, in such cases the computational cost of the experiment becomes very high, although the regression accuracy does not seem to require such amount of data in the training set. Therefore, as general rule of thumb, a good compromise between computational time and performance could be to limit the training sample to about $10^5$ samples.

In addition, our model MLPQNA has been tested in a public photo-z contest (PHAT1, Hildebrandt et al. 2010, \cite{hildebrandt2010}, and Cavuoti et al. 2012, \cite{cavuoti2012}), resulting as one of the best interpolative methods. In another work (Brescia et al. 2014, \cite{brescia2014b}) we published a catalogue of photometric redshifts for the SDSS DR9 release, by comparing our prediction accuracy with other machine learning methods. More recently PhotoRaptor has been used by an independent group (Hoyle et al. 2014, \cite{hoyle2014}), that performed a regression feature analysis with SDSS DR10 galaxies by comparing our resource with random forest (AdaBoost, \cite{drucker1997}) and FANN artificial neural networks (Nissen 2003, \cite{nissen2003}).

\section{Perspective and conclusions}
\label{conclusions}
Driven by the advances in the digital detectors and computing technology, astronomy has become an immensely data-rich science. This exponential data avalanche continues.
It enables some exciting new science, but poses many non-trivial challenges that are common to many other data-driven fields.
Nowadays the technological evolution of astronomical instruments has been so fast to render physically impossible to move the data from their
original repositories. The real goal of science, namely data analysis and knowledge discovery, begins after all the data processing and data delivery through the archives. This
requires some powerful new approaches to data exploration and analysis, leading to knowledge discovery and understanding. This implies that, as it has always been
asked for but never implemented, we must be able to move the programs and not the data.
Therefore, the future of any data-driven service depends on the capability and possibility of moving the data mining applications to the data centers hosting
the data themselves. In such scenario, PhotoRApToR represents our test bench of a desktop application prototype capable to fulfill this concept.
As a final perspective, we want to address the still open problem to find an efficient, reliable and standard way to provide single photo-z errors in the case of interpolative methods. We have recently started to investigate such problem and intend to improve PhotoRaptor in the next future with such kind of a tool.

\begin{acknowledgements}
\noindent The authors wish to thank the anonymous referee for all very useful comments and suggestions.
\noindent MB wishes to thank the financial support of the 7th European Framework Programme for Research Grant FP7-SPACE-2013-1, \textit{VIALACTEA -
The Milky Way as a Star Formation Engine}.
\noindent The authors also wish to thank the financial support of Project F.A.R.O. III Tornata (University Federico II of Naples).
\noindent GL acknowledges financial contribution through the PRIN-MIUR 2012 \textit{Cosmology with the Euclid Space Mission}.\\
\end{acknowledgements}



\end{document}